Science Tools Corporation
Street: 613 85th Avenue, Oakland CA 94621
Post: 4200 Park Blvd. #151, Oakland CA 94602
510-567-9957, FAX: 510-567-9975
http://ScienceTools.com/
December 2002

# Computational Unification:
# a Vision for Connecting Researchers

The extent to which the benefits of science can be fully realized depends critically upon the quality of the connection between researchers themselves and between researchers and members of the public. We believe that it is now possible to improve these connections on a community-wide and even world-wide basis through the use of an appropriate information management system. In this paper we explore the concepts and challenges, and propose an architecture for the implementation of such a system.

While a simple table consisting of research topics, researchers, and contact information, published visibly, could be considered a solution, we posit that such a solution, implemented in this new millenium, misses an overwhelmingly important opportunity and falls dramatically short of the existing potential. We instead propose that what is needed is a robust system to interconnect researchers at a much deeper level, utilizing the information management capabilities of computer systems to good effect. **This system's core responsibility is to manage collection and dissemination of the meta-data which provides the enabling infrastructure necessary to permit research systems to be connected using the latest technologies available, and to facilitate collaborative interaction and access by the lay public.** For browsing purposes, the system need not distinguish between other researchers and the public at large. Special considerations of the system regarding researchers are focused on the means by which the system is populated with information.

One of the greatest challenges of assembling a most useful means of connecting established researchers is overcoming the cacophony of voices and individualism which inherently exists. With the ubiquity of computing technology in research work today, it would seem natural to use computers as a means to bring harmony, but in scientific computing we find a plethora of solutions to what appear to be common problems. Each researcher has their own favorite data types and data-type hierarchies, data manipulation and visualization tools, system architectures and research paradigms. Given tight budgets, it is hard for individual researchers to see the benefits to them of creating systems which not only do their work but also connect them to others. And, often enough, researchers disagree with the technical arguments of their peers causing endless arguments and time-consuming battles over who's right whenever they endeavor to create computing systems for their disciplines. In sum, researchers are well-educated, intelligent people who have, with few exceptions, chosen to devote their lives to a discipline other than computer science; the information systems they create are tailored to specific problems and are in and of themselves not designed to be adapted to solve the problems of others, much less the more general problem of unifying a whole community, leaving as unthinkable and hardly imaginable a system which can unify scientific research as a whole. Yet increasingly it is becoming clear that this is precisely the need.

---

*"One of the greatest visions for science is a computational unification in which every researcher can interact with all other researchers through use of their own research system."*

---

To address this challenge and create a computer based information system which binds disparate researchers in a cohesive, unifying system, **the system must embrace the fundamental diversity as a feature.** As an example, a user of such a system who is interested in some particular type of scientific data may well wish to find:

- which sites and which specific computer systems have data collections of that type
- what scientific functions and processes operate against the type
- to what logical groupings that data type may belong
- what types of data are derived from that type and what processes perform that derivation
- which tools (software packages) are available to manipulate or visualize such data and which are the favorites of specific researchers
- what white-papers and web sites have been written about that data type
- who are the individuals responsible for any and all of the above

The answers to these questions, and many others, are stored in meta-data, without which the entire enterprise is not possible. Yet, most systems cannot answer these questions because they were not designed to unify a community and too many details are left as assumptions and presumptions. Discipline-specific, tailored solutions tend to fall short because of lack of consensus and cases where individual researchers may not conform to the ideas of peers in their field; a competent system must handle their data and meta-data with the same aplomb. The key is to have the right meta-data, organized using the right abstractions and using the best data storage technologies presently available.

**An existing system:** The critical need to provide a Community-Oriented approach was envisaged by the UC Berkeley led Sequoia 2000 project team, under Professor Michael Stonebraker, as an alternative to the Hughes EOS-DIS strategy; Emphasis was given to creating a high-performance earth-science system, fostering collaborative efforts with distributed processing, sophisticated searching, and strong scientific-defensibility features, with an initial focus on a single system to handle the needs of geology, hydrology, oceanography, and climate modeling. A prototype was built by the BigSur project team, and additional diverse data-sets were included to address end-user needs such as those of the State of California's Resources Agency. The technology was first demonstrated in the spring of '95. In the interim the technology was commercialized in 1997, has undergone further development by Science Tools corporation, and is performing production work at various research centers today. We call this the *BigSur System* (or just *BigSur*). What we are proposing here is at least one core installation of *BigSur* devoted to the purposes discussed in this white paper, working in conjunction with other *BigSur* installations or other scientific information systems which serve other specific needs whenever appropriate.

The most recent *Evolution* of *BigSur* pushes the collaborative paradigm even further, adding the concepts of research sites (not just individual computers) and the "publishing" of materials between sites. The *BigSur System* has a number of important attributes critical to the success of this work, among them:

> **Adaptive-Orientation:** An *Adaptive-Orientation* converts diversity into an asset to be exploited by learning the type descriptions and features of each researcher's strategies and paradigms. The system can also learn the specific processes and tools of each existing system in a direct way. **The goal is not to replace existing systems, but to harness them as tools.** For the individual researcher, an *Adaptive-Orientation* permits the system to learn the ways of the researcher rather than force the researcher to learn the ways of the system. And the system does not chase every new advancement in data access methods, such as new HDF formats, but rather it "learns" (or absorbs) those favored by the researcher as they are introduced. The system does not embed technology to manipulate or interpret data formats, rather, it records what data types exist, which objects are of those types, and what tools exist to handle them.

---

*"These features not only enable research collaboration on a scale never previously envisaged, they also enable sharing and dissemination of scientific knowledge to the public at large with a sophistication unparalleled in history."*

---

> **Progressive-Utilization:** *Progressive-Utilization,* **permits system implementers to pick and choose features as desired.** *Progressive-Utilization* extends flexibility by focusing on what the researcher finds appropriate, while minimizing requisite burdens. While the *BigSur System* was designed to manage Earth-system data from satellite to end-user desktop, *Progressive-Utilization* permits it to act as simply an electronic notebook, or just a distributed processing system, if that's all that is desired. In fact, the system does not even have to be installed at any given research facility in order to track and provide access to work performed there.
>
> **Database-Centrism:** Of primary interest to those wishing to create a computer based research information system are issues of meta-data management. Without meta-data, finding and sharing data is not possible; to facilitate fast, reliable, easy and ubiquitous meta-data access, a competent database system is a requisite core technology. The most capable system must take advantage of the fastest, most flexible and modern data management technology available, with as little demand for programming services as possible and a maximum ability to accept data-requests from applications of all varieties. Therefore, the system should be *database-centric*. *BigSur* is *database-centric* and a**ll meta-data is accessible through SQL to virtually all application programming environments.**

**Meta-Data Management:** Researchers create meta-data as they perform their work, such as the type-hierarchies and functional processes in which research methods and paradigms are implemented. Researchers also create white-papers, web pages, and other forms of textual or graphic presentation of findings. These meta-data fall into four categories: those which are "static", such as who researchers are, those which relate to particular theories and research themes, those which relate directly to individual data-objects, and those which are embedded as assumptions and presumptions into the information systems and application programs used. Exposing these data to others is vital. Because of the plethora of research data-types, tools, et cetera, within the community and the changing nature of these features over time, the system must be flexible, *Community* and *Adaptively-Oriented* and capable of *learning*. Thus, *BigSur* does not try to embed knowledge of these features, but rather learns (or absorbs) those favored by the researcher, and is free to pick up new advancements, such as the HDF5 data-type. **What are assumed relationships in other systems are explicitly stated in *BigSur*.** For example, the type representation of a "REGIS Aerial Photograph" is not assumed or implied to be a GIF file. Instead, if that is desired, the system is taught by making an entry associating these object types. Thus an object is properly represented as a "REGIS Aerial Photograph" and a GIF, both at the same time. (In addition, storage types and higher-level type semantics are managed separately.)

By applying our *Adaptive-Orientation* to all aspects of meta-data, and by not locking the researcher into predefined or rigid hierarchies, the system has deeply intimate knowledge and record of most aspects of scientific inquiry in its purview. Notable among these are lineage details about each individual scientific object, all the types and collections to which each belongs, the processes that created them and the parental input objects to those processes. Site-specific information is fully integrated with both data-objects and scientific processes, so object lineage can be maintained when a researcher uses as input data from a collaborator at another site. This permits strong scientific defensibility of conclusions reached through collaborative efforts throughout the community, regardless of where particular data-sets may be stored or meta-data managed.

***BigSur* can bring together and manage just the right kind of meta-data and make a reality the promise of a cohesive scientific research information system.**

**High-Performance, Distributed, Scaleable Architecture:** Site-independence of data-objects from meta-data provides a natural ability to cooperate in a distributed environment, and provide for sharing of workload between systems. ***BigSur's Distributed Processing System* permits Scientific functions to be performed on any system in the network, as desired, with both process-level and system-level controls for operations staff to manage workload.** At the present time, *BigSur* supports five different database systems: Oracle, Informix, Sybase, DB2, and Postgres. Use of modern ORDBMSs brings the benefit of robust tools, from many vendors and from the Open Source community, which can perform ancillary functions, aid in tailored application development, and so forth.

From an implementation perspective, more information is available for sharing and community-unifying purposes when scientific work includes elements which populate the information system automatically as it is conducted; There are many ways in which the system may be populated with data and meta-data starting with the automation of processing from within the system itself, and including technologies such as XML parsers of web-sites, email notification systems, FTP repositories, and so forth. Indeed, *BigSur* has been populated in precisely these ways in existing implementations. And a very significant contribution may be made "by hand", with a human being entering information in to the system on an as-needed basis. Some simple tools for this purpose may be made available so researchers may describe their work themselves, eliminating a human bottle-neck. As the installation grows, the distributed features may be used to divide workload and reduce resource loading such as network bandwidth. Specific disciplines may find the system of more interest than others and they may wish to have their own installation into which the community puts more than just superficial descriptions of their work - they may store actual utility programs, sample data-sets, and other introductory elements on the site. Some may decide that they like the system and wish to use it to help enable their own research, for example by using the *Distributed Processing System* as a means to automate workflow processing.

One of the greatest visions for science is a computational unification in which every researcher can interact with all other researchers through use of their own research system. The system we propose has all the right elements to do that: The capacity to describe the work of every researcher in their own terms, the ability to manage every type of data object and the functional processes that operate against them, and the ability to automate this processing. With some type conversions to perform transformations of data from one form to another - automated by the system - the system can join together the work of researchers from disparate disciplines. The system has full information about the associative relationships between scientific elements and so it "knows" the paradigms of each researcher. **It never forgets, so as people change, knowledge of how things are done is not lost.** These features not only enable research collaboration on a scale never previously envisaged, they also enable sharing and dissemination of scientific knowledge to the public at large with a sophistication unparalleled in history.

**Contacts:** Richard Troy: RTroy@ScienceTools.com, Olga Kingrey: Olga@ScienceTools.com